\documentclass{svjour3}

\usepackage{graphics}
\usepackage{amssymb}
\usepackage{bm}% bold math
\usepackage{natbib}
\usepackage{fancyhdr}

\usepackage{latexsym}

\usepackage[abs]{overpic}

\usepackage{amsfonts}

\usepackage{amsmath}

\usepackage{mathrsfs}

\usepackage{graphicx}

\newcommand{\pp}[1]{\phantom{#1}}
\newcommand{\be}{\begin{eqnarray}}
\newcommand{\ee}{\end{eqnarray}}

\begin{document}

\title{Response to the comment by Lambare}

\author{Marek Czachor}

\institute{Katedra Fizyki Teoretycznej i Informatyki Kwantowej,
Politechnika Gda\'nska, 80-233 Gda\'nsk, Poland}
\maketitle

\begin{abstract}
Contrary to what Lambare assumes, in non-Newtonian calculus (a calculus based on non-Diophantine arithmetic) an integral is typically given by a nonlinear map. This is the technical reason why all the standard proofs of Bell-type inequalities fail if non-Newtonian hidden variables are taken into account. From the non-Newtonian perspective, Bell's inequality is a property of a limited and unphysical class of hidden-variable models. An explicit counterexample to Bell's theorem can be easily constructed.
\end{abstract}

\keywords{Bell inequality   \and non-Diophantine arithmetics \and non-Newtonian calculus}

\section{}

Lambare in his comment \citep{L} on \citep{C} assumes that a linear combination of averages (6) implies a single average (7). This step would have  been justified if the integral that defines the averages had been given by a linear map. However, in non-Newtonian calculus, and this is the formalism one employs in the context of non-Diophantine arithmetic, an integral is linear only with respect to a very specific addition or multiplication, namely the ones implied by the non-Diophantine arithmetic in question. The same mathematical property occurs in fuzzy integration \citep{fuzzy}. The remark applies to all the standard proofs of Bell-type inequalities.

\section{}

Let me now briefly describe the non-Newtonian notion of an integral \citep{GK}, and then use it to construct a local hidden-variable model of two-electron singlet-state probabilities. For a more detailed analysis the readers are referred to the preprint \citep{C2}.

\begin{figure}
\includegraphics[width=8 cm]{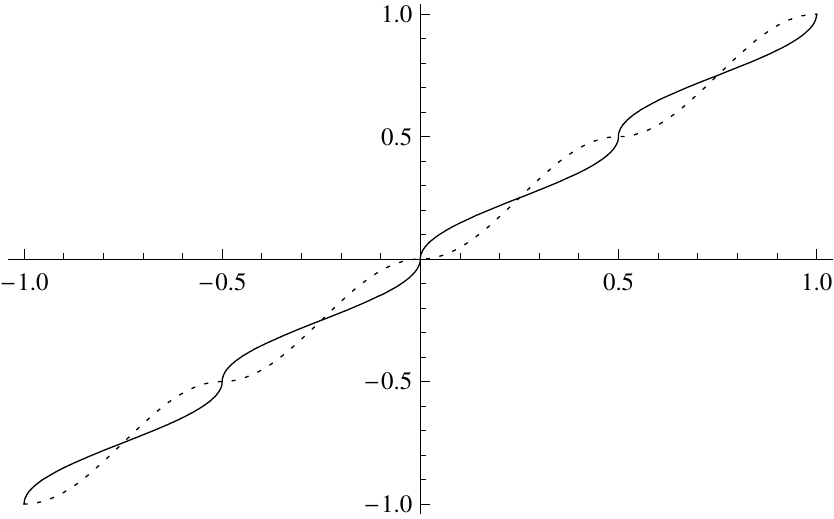}
\caption{One-to-one $f:\mathbb{R}\to \mathbb{R}$  (full) and its inverse $f^{-1}$ (dotted) defined by  (\ref{f})  and (\ref{f^-1}).}
\label{Fig2}
\end{figure}

Assume the set of hidden variables $\mathbb{X}$ has the same cardinality as the continuum $\mathbb{R}$. Accordingly, there exists a one-to-one map $f:\mathbb{X}\to \mathbb{R}$ which defines the arithmetic in $\mathbb{X}$,
\be
x\oplus y &=& f^{-1}\big(f(x)+f(y)\big),\label{ar1}\\
x\ominus y &=& f^{-1}\big(f(x)-f(y)\big),\label{ar2}\\
x\odot y &=& f^{-1}\big(f(x)\cdot f(y)\big),\label{ar3}\\
x\oslash y &=& f^{-1}\big(f(x)/f(y)\big).\label{ar4}
\ee
To make a long story short, take $\mathbb{X}=\mathbb{R}$, and
(Fig.~\ref{Fig2}),
\be
f^{-1}(x) &=&\frac{n}{2}+ \frac{1}{2}\sin^2\pi \left(x-\frac{n}{2}\right), \label{f^-1}\\
f(x) &=& \frac{n}{2}+\frac{1}{\pi}\arcsin\sqrt{2x -n}, \label{f}\\
&\pp=&
\textrm{for $\frac{n}{2}\le x\le \frac{n+1}{2}$, $n\in\mathbb{Z}$}.
\ee
Notice that $f(k/4)=f^{-1}(k/4)=k/4$ for any integer $k$. 
A function  $a:\mathbb{X}\to \mathbb{X}$ defines a new function $\tilde a:\mathbb{R}\to \mathbb{R}$ such that the diagram 
\be
\begin{array}{rcl}
\mathbb{X}                & \stackrel{a}{\longrightarrow}       & \mathbb{X}               \\
f{\Big\downarrow}   &                                     & {\Big\downarrow}f   \\
\mathbb{R}                & \stackrel{\tilde a}{\longrightarrow}   & \mathbb{R}
\end{array}\label{diagram}
\ee
is commutative. 
The derivative of $a$ is defined as \citep{GK}
\be
\frac{{\rm D}a(x)}{{\rm D}x}
&=&
\lim_{\delta\to 0}\Big(
a(x\oplus \delta)\ominus a(x)\Big)\oslash \delta.
\label{DA}
\ee
Once we have the derivatives we define a non-Newtonian  integral of  $a$ by the fundamental theorem of non-Newtonian calculus \citep{GK},
\be
\int_{x_1}^{x_2} a(x){\textrm D}x
&=&
f^{-1}
\left(
\int_{f(x_1)}^{f(x_2)}\tilde a(r){\textrm d}r
\right)\label{integr}
\ee
i.e. in terms of the Newtonian  integral of $\tilde a$. It is now clear why, in general, 
\be
\int_{x_1}^{x_2} \big(a(x)+b(x)\big){\textrm D}x
\neq
\int_{x_1}^{x_2} a(x){\textrm D}x
+
\int_{x_1}^{x_2} b(x){\textrm D}x,
\ee
so that standard Bell-type inequalities cannot be proved (unless $f(x)$ is linear). Instead, what one gets is a non-Diophantine linearity,
\be
\int_{x_1}^{x_2} \big(a(x)\oplus b(x)\big){\textrm D}x
=
\int_{x_1}^{x_2} a(x){\textrm D}x
\oplus
\int_{x_1}^{x_2} b(x){\textrm D}x.
\ee
Now denote $x'=f^{-1}(x)$. In order to construct a hidden-variable model of singlet-state correlations we begin with non-Newtonian representation of probability density. The diagram, 
\be
\begin{array}{rcl}
\mathbb{X}                & \stackrel{\rho}{\longrightarrow}       & \mathbb{X}               \\
f{\Big\downarrow}   &                                     & {\Big\downarrow}f   \\
\mathbb{R}                & \stackrel{\tilde \rho}{\longrightarrow}   & \mathbb{R}
\end{array},
\ee
where
\be
1 =
\int_{0}^{(2\pi)'}\rho(x){\rm D}x
=
f^{-1}\left(
\int_{0}^{2\pi}\tilde\rho(r){\rm d}r
\right)=
f^{-1}(1)
\nonumber
\ee
implies $\tilde\rho(r)=1/(2\pi)$, so we arrive at
\be
\rho(x) &=& f^{-1}\circ\tilde\rho\circ f(x)=f^{-1}\big(1/(2\pi)\big)\approx 0.114924.
\ee
 Then \citep{C2}, for $0\le \beta-\alpha\le \pi$, 
\be
\int_{\alpha'}^{\beta'}\rho(x){\rm D}x
=
f^{-1}\left(
\frac{\beta-\alpha}{2\pi}
\right)=
\frac{1}{2}\sin^2\frac{\beta-\alpha}{2}
.\label{sin}
\ee
One can analogously define all the remaining joint probabilities typical of two-electron singlet states. Moreover, all these probabilities have the form typical of local-realistic Clauser-Horne hidden-variable models, namely
\be
P_{++}(\beta-\alpha)
&=&
\frac{1}{2}\sin^2\frac{\beta-\alpha}{2}
=
\int_{\alpha'}^{\beta'} \rho(\lambda){\rm D}\lambda\label{rho1}\\
&=&
\int_{0}^{(2\pi)'} \chi_{\alpha +}^1(\lambda)\odot \chi_{\beta +}^2(\lambda)\odot
\rho(\lambda){\rm D}\lambda,\label{rho1'}\\
P_{+-}(\beta-\alpha)
&=&
\frac{1}{2}\cos^2\frac{\beta-\alpha}{2}
=
\int_{\beta'}^{\alpha'\oplus\pi'} \rho(\lambda){\rm D}\lambda\label{rho2}\\
&=&
\int_{0}^{(2\pi)'} \chi_{\alpha +}^1(\lambda)\odot \chi_{\beta -}^2(\lambda)\odot
\rho(\lambda){\rm D}\lambda,\label{rho2'}\\
P_{--}(\beta-\alpha)
&=&
\frac{1}{2}\sin^2\frac{\beta-\alpha}{2}
=
\int_{\alpha'\oplus\pi'}^{\beta'\oplus\pi'} \rho(\lambda){\rm D}\lambda\label{rho3}\\
&=&
\int_{0}^{(2\pi)'} \chi_{\alpha -}^1(\lambda)\odot \chi_{\beta -}^2(\lambda)\odot
\rho(\lambda){\rm D}\lambda,\label{rho3'}\\
P_{-+}(\beta-\alpha)
&=&
\frac{1}{2}\cos^2\frac{\beta-\alpha}{2}
=
\int_{\beta'\oplus\pi'}^{\alpha'\oplus(2\pi)'} \rho(\lambda){\rm D}\lambda\label{rho4}\\
&=&
\int_{0}^{(2\pi)'} \chi_{\alpha -}^1(\lambda)\odot \chi_{\beta +}^2(\lambda)\odot
\rho(\lambda){\rm D}\lambda.\label{rho4'}
\ee
The model we consider is deterministic, local-realistic, rotationally invariant, observers have free will, detectors are perfect,  so is free of all the canonical loopholes discussed in the literature. 
It satisfies all the desiderata of the original Bell construction, and involves Einstein-Podolsky-Rosen elements of reality. 
Yet, it explicitly contradicts the conclusions of Bell (and Lambare).

\end{document}